\PassOptionsToPackage{numbers,sort&compress}{natbib}  
\documentclass[preprintnumbers,amsmath,amssymb,prd, notitlepage,nofootinbib,twocolumn, superscriptaddress]{revtex4}

\usepackage{graphicx}
\usepackage{dcolumn}
\usepackage{bm}
\usepackage{subfigure}
\usepackage{wasysym}
\usepackage{verbatim}
\usepackage{color}
\usepackage{amsmath}

\newcommand{\be}{\begin{equation}}
\newcommand{\ee}{\end{equation}}

\newcommand{\rmn}{\mathrm}

\newcommand{\ba}{\begin{eqnarray}}                                  
\newcommand{\ea}{\end{eqnarray}}                                    
\newcommand{\nn}{\nonumber}                                          
\newcommand{\barr}{\begin{array}}                                    
\newcommand{\earr}{\end{array}}

\def\n{{\bf \hat{n}}}

\def\k{{\bf k}}
\def\vrec{\hat{v}_{\mathrm{rec}}}

\begin{document}

\title{Cosmology with kSZ: breaking the optical depth degeneracy with Fast
Radio Bursts}

\author{Mathew~S.~Madhavacheril}
\affiliation{Princeton University, Department of Astrophysical Sciences, Princeton NJ 08540, USA}
\author{Nicholas Battaglia}
\affiliation{Department of Astronomy, Cornell University, Ithaca, NY, 14853, USA}
\affiliation{Center for Computational Astrophysics, Flatiron Institute, 162 5th Avenue, 10010, New York, NY, USA}
\author{Kendrick M. Smith}
\affiliation{Perimeter Institute for Theoretical Physics, Waterloo, ON N2L 2Y5, Canada}
\author{Jonathan L. Sievers}
\affiliation{Department of Physics, McGill University, 3600 Rue University, Montreal QC Canada}
\affiliation{School of Chemistry and Physics, University of KwaZulu-Natal, Westville Campus, Durban South Africa}
\date{\today}

\begin{abstract}
The small-scale cosmic microwave background (CMB) is dominated by anisotropies
from the kinematic Sunyaev-Zeldovich (kSZ) effect, and upcoming experiments will measure it very
precisely, but the optical depth degeneracy limits the cosmological information
that can be extracted.  At the same time, fast radio bursts (FRBs) are an exciting new frontier for
astrophysics, but their usefulness as cosmological probes is currently unclear.
We show that FRBs are uniquely suited for breaking the kSZ optical depth
degeneracy.  This opens up new possibilities for constraining cosmology with the
kSZ effect, and new cosmological applications for FRBs.
\end{abstract}

\maketitle

\section{Introduction}

As photons from the cosmic microwave background (CMB) travel through the
Universe, a small fraction interact with free electrons. The kinematic Sunyaev-Zeldovich (kSZ) effect
\cite{SunyaevZeldovichKinematic} is the
result of CMB photons Compton scattering off free electrons that have non-zero
peculiar velocities with respect to the CMB rest frame, which lead to additional
anisotropies in the observed CMB radiation. As a result, we observe a small
shift in the CMB temperature in the direction of those free electrons. This
shift is proportional to the integrated momentum along the line-of-sight. Thus,
kSZ measurements are potentially powerful observational probes of the peculiar
velocities of systems of ionized gas that trace the total distribution of matter
\citep[e.g.,][]{Ferreira1999,AGP2001,HM2006,BK2007}. Since the small-scale CMB
is dominated by kSZ fluctuations and upcoming CMB surveys will measure it very
precisely \cite{SO,SmithKSZ2018}, mapping the peculiar
velocity distribution of the Universe with kSZ will provide competitive constraints on
primordial non-Gaussianity \cite{kszFnl}.  Velocities probe the
cosmological growth rate, which can allow further constraints on modified gravity models, the dark energy equation of
state, and the sum of neutrino masses  \citep[e.g.,][]{Dedeo2005,HM2006,BK2008,KS2013,Mueller2015a,Mueller2015b,ALBF2016,SO,kszFnl}. 

The cosmological growth rate measured through the kSZ effect is however
perfectly degenerate with the optical depth of galaxies or clusters
\citep[e.g.,][]{Battaglia2016,Flender2017} leading to an overall uncertainty in
the inferred amplitude of the growth rate. This degeneracy with the optical
depth is the limiting systematic uncertainty for measurements of the
cosmological growth rate from kSZ tomography
\citep[e.g.,][]{Mueller2015a,ALBF2016,SmithKSZ2018}.

Recent detections of multiple Fast Radio Burst Sources (FRBs\footnote{Hereafter,
  FRBs refers to the sources, not the bursts.}) along with theoretical models
strongly suggest that there exist transient radio events originating (possibly)
from energetic events at cosmological redshifts that are detectable with a rate
greater than one per day \citep[e.g.,][]{CHIMEFRB2017,ATelFRB2018,ShannonNature2018}. With future
upgrades and outrigger stations, instruments like HIRAX \citep{HIRAX} should
localize of order 10 FRBs per day with sub-arcsecond accuracy which will enable
one to acquire redshifts \citep{Tendulkar2017}. Plasma along the line of sight
delays the FRBs in a frequency-dependent manner, with the delay in seconds
approximately equal to $4.15\times10^{-3} \rm{DM}/\nu_{\rm{GHz}}^2$ where DM is
the {\it dispersion measure}, in $\rm{pc}/\rm{cm}^3$, and is equivalent to the
optical depth $\tau$ due to Compton scattering: $\mathrm{DM}= (4.87 \times 10^5 \,\, \mathrm{pc} \,\, \mathrm{cm}^{-3}) \, \tau$.
Radio telescopes measure the DMs associated with these events quite precisely
(typical measurement accuracies are 0.1\%), which receive contributions from the
host galaxy, the Milky Way, and any intervening free electrons
\citep[e.g.,][]{Dolag2015}. The third of these contributions is of great
interest to the extragalactic and cosmological communities. With the promise of
thousands of FRBs in the future, theoretical ideas and forecasts have been
published regarding measuring the baryon content in the Intergalatic Medium
\citep[IGM, e.g.,][]{McQuinn2014} and the Circumgalactic Medium \citep[CGM,
  e.g.,][]{Ravi2018}, regarding constraining the reionization epoch \cite{FRBTauCMB}, and regarding measuring 3D clustering of free electrons \cite{Masui2015} to name a few.

In this work, we propose to directly measure the galaxy optical depth
through the contribution to FRB DMs from scattering off of
intervening free electrons, using the cross-correlation between the galaxy
sample used in the kSZ measurement and a map of FRB dispersion measures. This
cross-correlation can be directly interpreted as the galaxy optical depth as it
is measuring the galaxy-electron power spectrum $P_{ge}(k)$, thus breaking the
optical depth degeneracy and allowing for sub-percent constraints on the growth
rate. We focus on the information on the cosmological growth rate that we can
extract with thousands or more of localized FRB measurements in combination with
kSZ measurements made by upcoming CMB and galaxy surveys. We note that a recent
paper \cite{FRBCosmo} investigated the possibility of using FRBs for
cosmological tests, but found no interesting applications other than
constraining the ionized gas distribution. We show in this work that
constraining ionized gas (specifically, the galaxy-electron correlation) with FRBs
enables cosmological applications of the kSZ effect.

\section{The galaxy-electron spectrum measured with FRBs}

The dispersion measure (DM) along a line of sight should be correlated with the density of foreground galaxies in that direction, since some of the electron fluctuations contributing to the DM originate from those galaxies.
We are thus interested in cross correlating foreground galaxies with a map of
DMs (not spatial locations) from FRBs. Note that this does not require FRBs to
be clustered or for them to have redshift overlap with the galaxies. They
instead act as a backlight for the free electrons in these galaxies, like quasars act for neutral hydrogen.
The FRBs need to be localized with redshift information sufficient to inform whether or not the FRB in any given FRB-galaxy pair is behind the galaxy.

Because the DM is an integrated quantity along the line of sight, it is
convenient to do the forecast using 2-d fields (not 3-d).  For this preliminary
investigation into the feasibility of using FRBs for cosmology, we work with a
simplified geometry. We consider a thin shell of foreground galaxies,
specifically a sample with a mean redshift of 0.75, redshift shell width of
 0.3 and number density of $\sim 1.7\times 10^{-4} \,\, \rmn{Mpc}^{-3}$ expected to be
 provided by surveys like the Dark Energy Spectroscopic Instrument (DESI) \cite{DESIScience}. All the FRBs are assumed to lie in a thin background shell
centered at $z=1$.  In this thin-shell geometry, we can treat all fields in sight as 2-d fields.

Let $(\chi_g- \Delta\chi_g/2, \chi_g + \Delta\chi_g/2)$ be the comoving distance interval spanned by
the foreground galaxies, and let $(\chi_f- \Delta\chi_f/2, \chi_f + \Delta\chi_f/2)$ be the comoving
distance interval spanned by the background FRBs.  We will use the notation $(\cdot)_g$
to mean ``evaluated at the redshift of the galaxies'', e.g.~$z_g$ is the galaxy redshift.

We assume that the separation between the foreground and background shells is large enough
that there are no spatial correlations between foreground galaxies and the spatial locations
(or the host DMs) of background FRBs.  Thus any galaxy-DM correlations can be attributed
to correlations between the galaxies and the electrons along the line of sight in those galaxies.

\begin{figure}[tbh]
  \includegraphics[width=0.95\columnwidth]{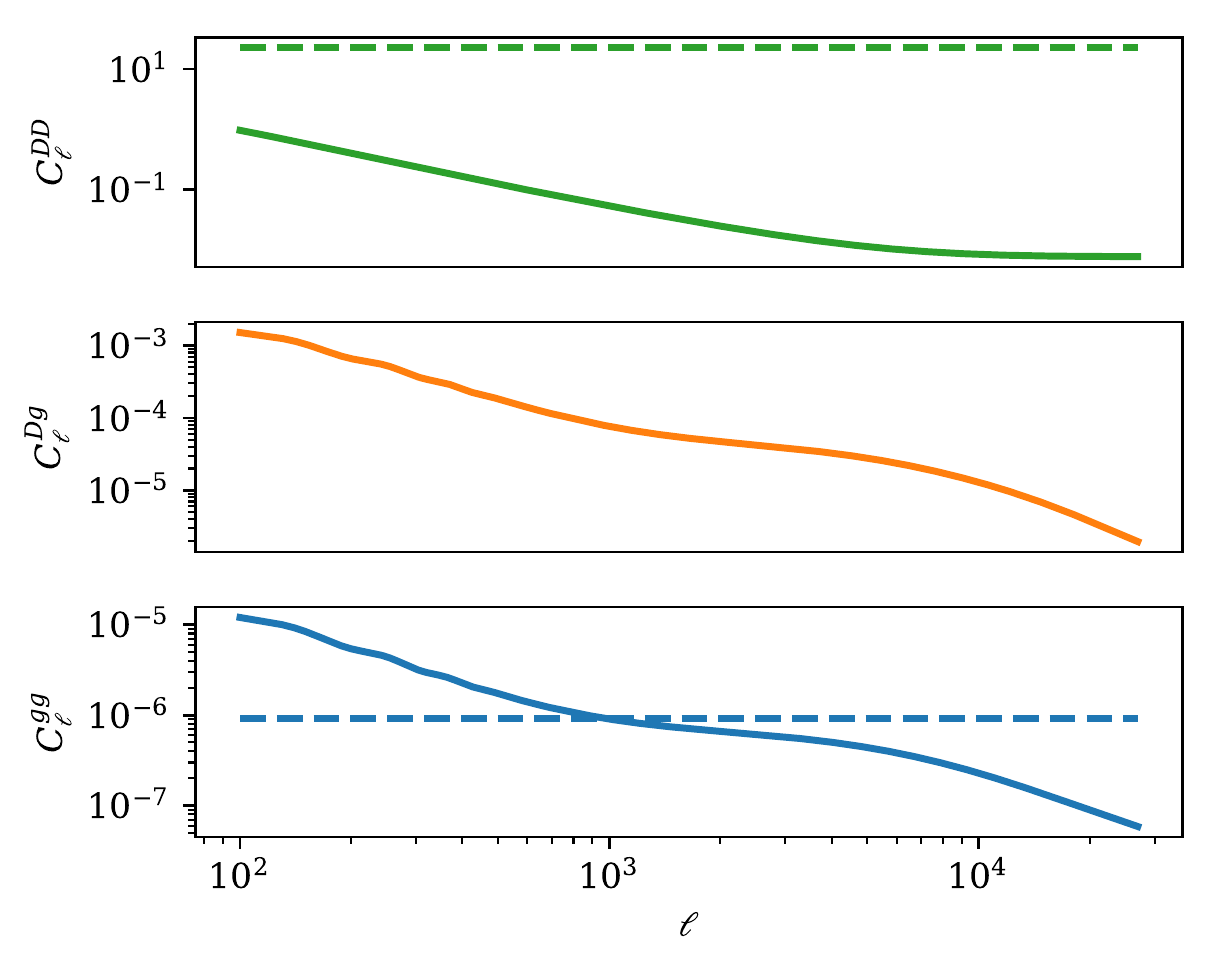}
\caption{Power spectra of the FRB dispersion measure and galaxy density fields calculated under the thin-shell approximation. The solid green, orange, and blue lines show the FRB DM auto power spectrum (in $\left(\frac{\mathrm{pc}}{\mathrm{cm}^3}\right)^2$), the galaxy-DM cross power spectrum (in $\frac{\mathrm{pc}}{\mathrm{cm}^3}$), and the galaxy auto power spectrum (dimensionless), respectively. The blue dashed line shows the shot noise per mode in the DESI galaxy survey. The green dashed line shows the effective noise per mode in the FRB DM field for a DM RMS scatter of 300 $\frac{\mathrm{pc}}{\mathrm{cm}^3}$ and total number of FRBs of 10,000.}
\label{fig:clggs}
\end{figure}

The line-of-sight integral for the dispersion measure is (see e.g. \cite{Ioka2003}):
\be
D(\n) = n_{e0} \int_0^{\chi_f} d\chi \, (1+z) (1 + \delta_e(\n,z)),
\ee
where $\n$ is the line of sight direction, $n_{e0}$ is the mean number density
of free electrons at $z=0$, and $\chi$ is the comoving distance. In the Limber
approximation (equivalent to a small-angle approximation which is valid for the
scales we consider), the cross-correlation between the 2-d DM field, $D$, and the 2-d
galaxy overdensity field, $\delta_g$, is:
\be
C_l^{Dg} = n_{e0} \, \frac{1+z_g}{\chi_g^2} \, P_{ge}(k,z_g)_{k=l/\chi_g}  \label{eq:cldg_limber},
\ee
where the $C_l$ notation denotes angular power spectra at angular wavenumber or
multipole $l$, and $P_{ge}$ is the 3D galaxy-electron cross power spectrum as a
function of the magnitude $k$ of the 3D Fourier wavenumber ${\bf k}$. Our
proposed observable $C_l^{Dg}$ thus measures the power
spectrum $P_{ge}$, which is an important quantity that captures how the free electron
overdensity $\delta_e$ is correlated with the galaxy overdensity $\delta_g$. As
explained in \cite{SmithKSZ2018}, the very same power spectrum $P_{ge}$ is also measured
by kSZ tomography. However, for cosmological applications of kSZ, $P_{ge}$
appears in a nuisance parameter that multiplies the cosmologically informative
cross power spectrum of galaxies and the cosmic velocity field $P_{gv}$ \cite[e.g.,][]{SmithKSZ2018}. This
motivates our external measurement of $P_{ge}$ from FRB DMs.

To complete our forecast for the signal-to-noise-ratio (SNR) of $C_l^{Dg}$, we also need the associated auto power spectra (again making the Limber approximation):
\ba
C_l^{DD} &=& n_{e0}^2 \int_0^{\chi_f} d\chi \, \frac{(1+z)^2}{\chi^2} \, P_{ee}(k,z)_{k=l/\chi},  \\
C_l^{gg} &=& \frac{1}{\chi_g^2 (\Delta\chi_g)} P_{gg}(k,z_g)_{k=l/\chi_g}  \label{eq:clgg_limber},
\ea
where $P_{ee}$ and $P_{gg}$ are the electron and galaxy auto power spectra, respectively. 

\begin{figure*}[tbh]
  \includegraphics[width=1.9\columnwidth]{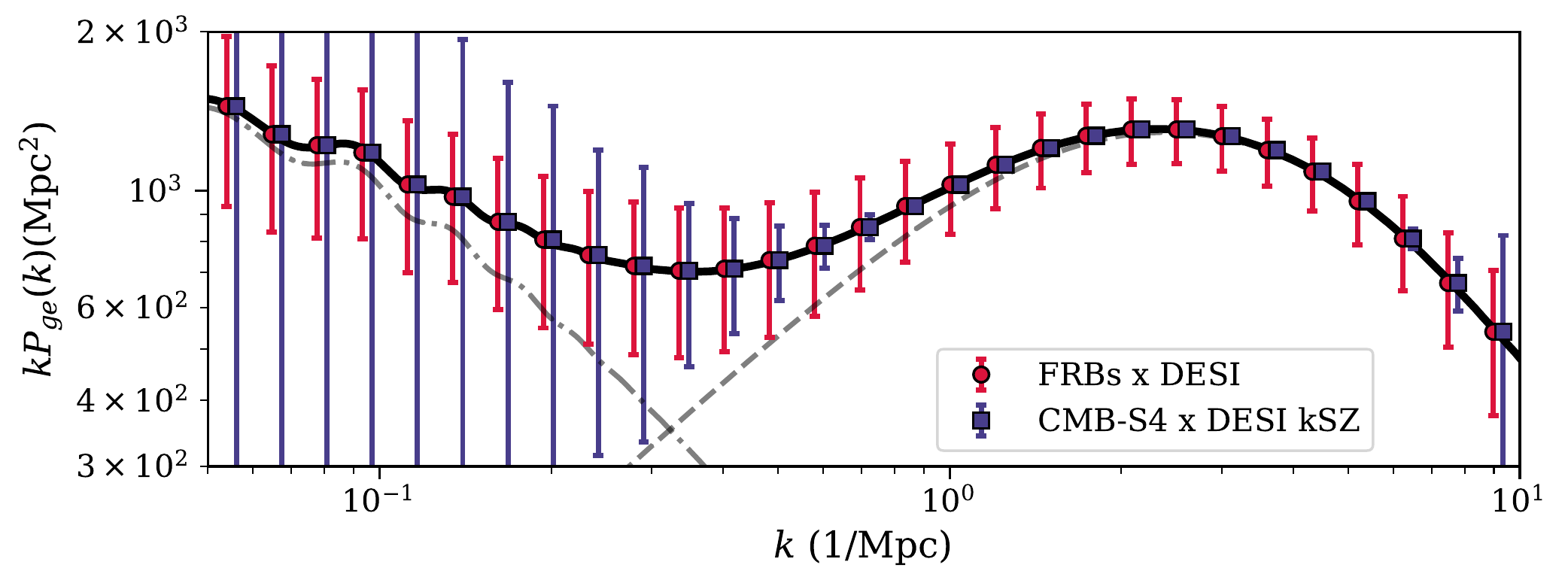}  
\caption{The cross-power-spectrum of galaxies and electrons as measured either
  through kSZ tomography with CMB-S4 and DESI (blue) with fixed cosmology, or
  through cross-correlation of dispersion measures of $10^4$ FRBs with DESI galaxies
  (red), where the RMS
  scatter of DMs is assumed to be 300 $\frac{\mathrm{pc}}{\mathrm{cm}^3}$. FRB
  DMs measure the power over a broad range of scales including the 2-halo regime
  (dot-dashed), while kSZ tomography provides an
  extremely tight measurement in the 1-halo dominated regime (dashed). Our lack of knowledge of the galaxy-electron power spectrum on these small scales limits our ability to use large-scale velocities from kSZ for cosmology. This degeneracy can be broken using the externally measured FRB cross-correlation.}
\label{fig:pge}
\end{figure*}

The small-scale power spectra above are calculated in the halo model following
\cite{SmithKSZ2018}, with contributions from clustering of electrons and
galaxies (the 2-halo term) and from the shape of the profiles of the electron
and galaxy distributions (the 1-halo term). The calculated 2-d power spectra are
shown in
Fig.~\ref{fig:clggs}. 
When we ``observe'' the 2-d DM field, $D$, with a discretely sampled catalog of FRBs,
there is an associated noise power spectrum $N_l^{DD}$ given by:
\be
N_l^{DD} = \frac{ \sigma_D^2 }{n_f^{2d}}.  \label{eq:nldd}
\ee
Here, $n_f^{2d}$ is the number density (per steradian) of FRBs, and $\sigma_D^2$ is the total variance of the DMs.
The latter is the sum of three contributions: intrinsic scatter in the FRB host's DM,
residual uncertainty in the DM of our galaxy, and a term $\int d^2l/(2\pi)^2 \, C_l^{DD}$ 
from electron fluctuations along the line of sight that not associated with
galaxies we are cross correlating with, the cosmological variance. We will not
worry about keeping track of these contributions separately, since the host
contribution is a free parameter anyway. Since the RMS scatter of the DMs
$\sigma_D$ is currently uncertain, we show forecasts for various plausible
values given current detections of FRBs. We chose the range to be from 100
pc/cm$^3$ to 1000 pc/cm$^3$. This range is motivated by empirical measurements
of the intrinsic DM of the host of the repeating FRB \citep{Chat2017}, which has
DM of $\le 324$ pc/cm$^3$ \citep{Tendulkar2017}. The cosmological DM RMS scatter is
of order 100-1000 pc/cm$^3$ from our halo model calculations and from simulations \citep{Dolag2015}. The DM of our galaxy varies dramatically depending on sky location, however models exists \citep[e.g.,][]{NE2001} to remove this contribution with only a small, uncorrelated residual left to contribute to the variance.

In terms of the above definitions, the total $S/N$ of the DM-galaxy cross power
is given by:
\be
\mbox{S/N}^2 = \Omega \int \frac{d^2l}{(2\pi)^2} \frac{ (C_l^{Dg})^2 }{(\mathcal{N}_l^{Dg})^2}, \label{eq:thin_shell_snr}
\ee
where
\be
(\mathcal{N}_l^{Dg})^2 = (C_l^{gg} + 1/n_g^{2d}) (C_l^{DD} + \sigma_D^2 / n_f^{2d}),
\ee
$n_g^{2d}$ is the number density of galaxies in the galaxy survey (per
steradian), and $\Omega$ is the angular size of the survey in steradians which accounts for the partial sky coverage fraction $f_{\rm{sky}}$ of the survey overlap through $\Omega=4\pi f_{\rm{sky}}$.

Using Eq.~\ref{eq:cldg_limber} and Eq.~\ref{eq:thin_shell_snr}, we can also obtain the uncertainty on the bandpowers of the
galaxy-electron power inferred from the DM-galaxy cross correlation (see
Appendix A for details),

\begin{multline}
\Delta P_{ge} = \frac{\chi_g}{n_{e0} (1+z_g)} \left( \Omega \int_{k_{\rm min}}^{k_{\rm max}} \frac{k \, dk}{2\pi}
    \frac{ 1 }{ (\mathcal{N}_l^{Dg})^2 } \right)_{l=k\chi_g}^{-1/2}  \label{eq:delta_pge_frb}
\end{multline}

\begin{figure}[tbh]
  \includegraphics[width=0.95\columnwidth]{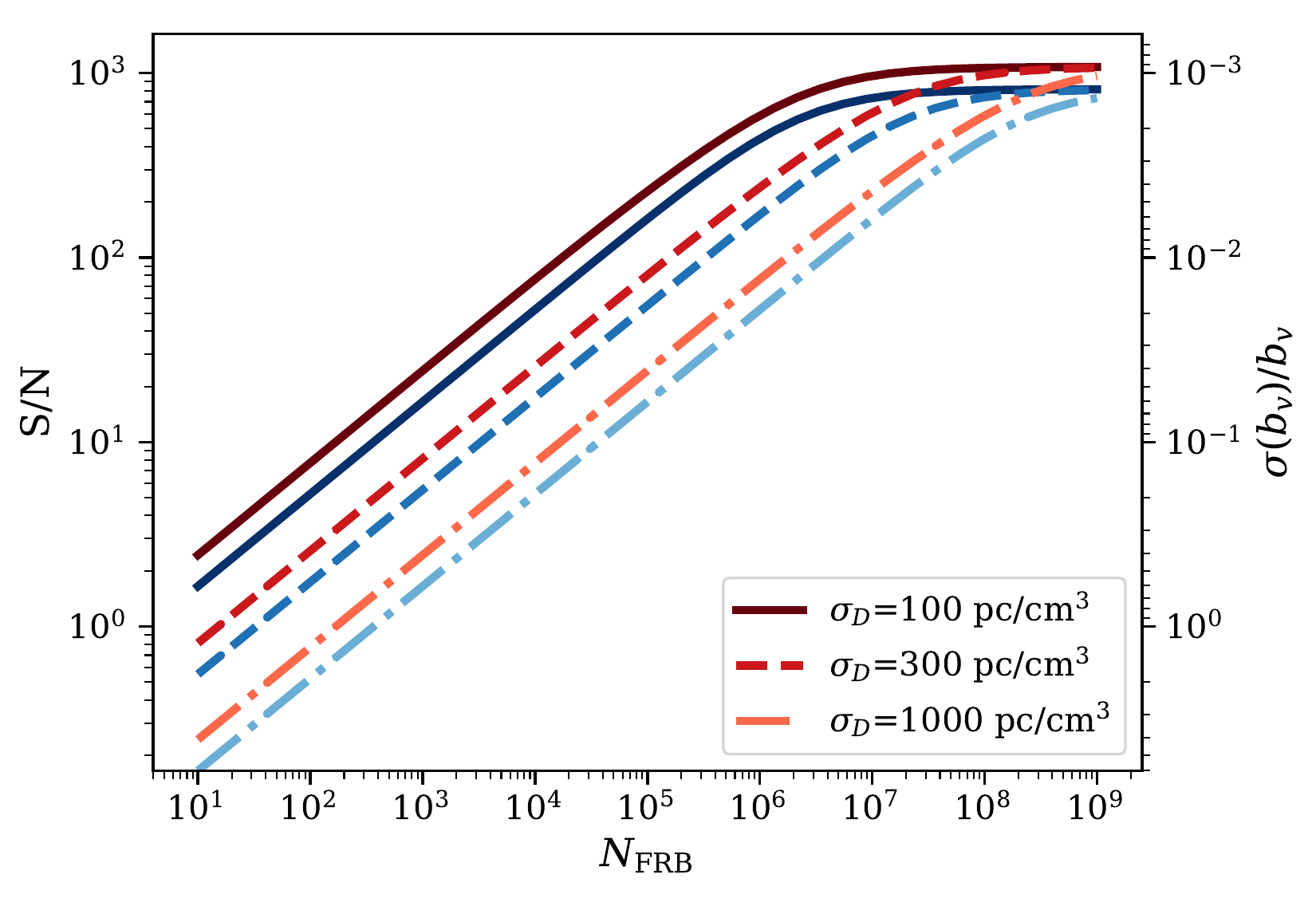}  
\caption{ The signal-to-noise ratio ($S/N$) of the cross-correlation of DMs from
  FRBs with the DESI galaxy survey (red, left vertical axis) and the closely
  related relative uncertainty on the velocity bias or equivalently the galaxy optical depth (blue, right vertical
  axis), as a function of the number of FRBs, $N_{\rm{FRB}}$, in the background of DESI galaxies (over an overlap $f_{\mathrm{sky}}=0.2$). The $S/N$ depends strongly on the currently poorly constrained RMS scatter of intrinsic DM of the FRB host galaxy, here shown for various plausible values. Sample variance in the free electron fluctuations causes the $S/N$ to saturate to $\approx 10^3$.} 
\label{fig:sn}
\end{figure}

In Fig.~\ref{fig:pge}, we show the galaxy-electron power spectrum along with the uncertainties on its bandpowers from a measurement made using the DESI galaxy
sample cross-correlated with $10^4$ FRB DMs, assuming the simplified geometry
described above and a DM RMS scatter of 300
$\frac{\mathrm{pc}}{\mathrm{cm}^3}$. For comparison, we also show the
uncertainties on the $P_{ge}$ from a kSZ tomography \cite{SmithKSZ2018}
measurement using the proposed CMB-S4 experiment \cite{CMBS4} and DESI, where we
have assumed that the factor that multiplies $P_{ge}$ and depends on the cosmologically informative power spectrum $P_{gv}$ has
been fixed to a fiducial cosmology. We see that FRB DMs measure $P_{ge}$ over a
broad range of scales, while as noted in \cite{SmithKSZ2018}, kSZ tomography
measures it very well only in a small range of scales in the 1-halo regime.

\section{The Cosmological connection}

The FRB-determined measurement of the small-scale cross-power-spectrum of galaxies and
electrons $P_{ge}(k)$ detailed in the previous section can be used to break a
degeneracy that limits the cosmological utility of kSZ tomography. Since the kSZ
effect arises from the Doppler shifting of CMB photons that Compton scatter off
free electrons with bulk radial velocities, the large scale cosmic velocity field
modulates the cross-power-spectrum of the CMB temperature and galaxy overdensity
field. This idea allows one to infer the large-scale cosmic velocity field from a
combination of the CMB temperature anisotropies as measured by a CMB survey and
the positions of galaxies as measured by a galaxy survey \cite{Deutsch2017,SmithKSZ2018} on small scales. However, this velocity
field can only be inferred up to an overall constant $b_v$ since the kSZ effect
is proportional to both the bulk radial velocity and the overdensity of free
electrons. This unknown constant $b_v$ is in fact an integral over precisely the
small-scale galaxy-electron power spectrum $P_{ge}(k)$ that can be measured with FRB
DMs. 

On large scales where linear theory is valid, the velocity reconstruction from
kSZ tomography is directly proportional to the cosmic growth rate $f(a)\approx \frac{d\mathrm{ln}D(a)}{d\mathrm{ln}a}$, where $D(a)$ is the growth factor for the matter spectrum that evolves as $P_{mm}(a)=D^2(a)P_{mm}(a=1)$
and $a$ is the expansion scale factor. Since the velocity reconstruction is
uncertain up to the amplitude $b_v$, in order to convert a kSZ
tomography measurement to cosmological information on
massive neutrinos, dark energy perturbations, modified gravity and other physics
that can affect the growth rate, one needs an external measurement of $b_v$, or
equivalently of $P_{ge}(k)$. This is the so-called ``optical depth
degeneracy''.\footnote{Hereafter, we will refer to the unknown quantity (whose
  priors we obtain from FRBs) as the `velocity bias', which can be loosely interchanged with `optical depth'. } To summarize, the program of constraining cosmology using linear theory with large-scale ($k<<0.1 \rm{Mpc}^{-1}$ ) velocities from kSZ requires knowledge of an integral of the galaxy-electron power spectrum over extremely non-linear small scales ($0.1 $ Mpc$^{-1} < k < 10 $ Mpc$^{-1}$). \footnote{Note however that scale-dependent effects, e.g. scale-dependent galaxy bias from primordial non-Gaussianity, can  be constrained extremely well \cite{Munchmeyer2018}.}

We have seen in the previous section that FRB DMs can provide this
external measurement of small-scale $P_{ge}(k)$. At the back-of-the-envelope level, a 1\% constraint on $P_{ge}(k)$
from FRBs (or equivalently $\rmn{S/N}=100\sigma$ on $C_{\ell}^{Dg}$)
could translate to a 1\% constraint on the velocity bias $b_v$. However, in
practice, the
velocity bias information in the FRB measurement is somewhat lower, because FRB DMs measure
$P_{ge}(k)$ over a broad range of scales while (due to the squeezed bispectrum
origin of the kSZ effect) the optical depth degeneracy is
sourced primarily by small scales in the ``1-halo'' regime. In Appendix B, we obtain the
1-sigma constraint $\sigma(b_v)$ from FRB DMs properly accounting for this.

In Fig.~\ref{fig:sn} we show both the raw SNR for the measurement of $C_{\ell}^{Dg}$ (or
$P_{ge}(k)$) using FRB DMs and the DESI galaxy survey (from Eq.~\ref{eq:thin_shell_snr}), and the closely related
relative uncertainty on the velocity bias calculated using Eq.~\ref{eq:sig_bv}. As expected
the SNR on the velocity bias is slightly lower. The SNR saturates at high FRB
number density when it becomes limited by the sample variance $C_l^{DD}$ in the
contribution to DMs from intervening free electrons.

We can now obtain constraints on the cosmic growth rate that incorporate prior
information on $b_v$ from FRBs. The large scale velocity field in linear theory inferred from kSZ tomography is now,
\be                                                                
\vrec(\k) = b_v \mu \frac{faH}{k} \delta_m(\k),
\ee  
where $\k$ is the 3-d wavevector, $\mu=k_r/k$ for radial component of the
wavevector $k_r$ (along the line of sight), $H$ is the Hubble constant at the
redshift of the galaxy sample and $\delta_m$ is the matter overdensity. The
velocity reconstruction is performed over small-wavelength modes $k_S$ in the
high-resolution CMB survey and the galaxy survey. The modes $k_S$ are limited to
$0.1 $ Mpc$^{-1} < k < 10 $ Mpc$^{-1}$.

We marginalize over $b_v$ for a fiducial value of $b_v=1$ but with the Gaussian
prior determined earlier that depends on the number of FRBs, $N_{\rm{FRB}}$.

\begin{figure}[tbh]
  \includegraphics[width=0.95\columnwidth]{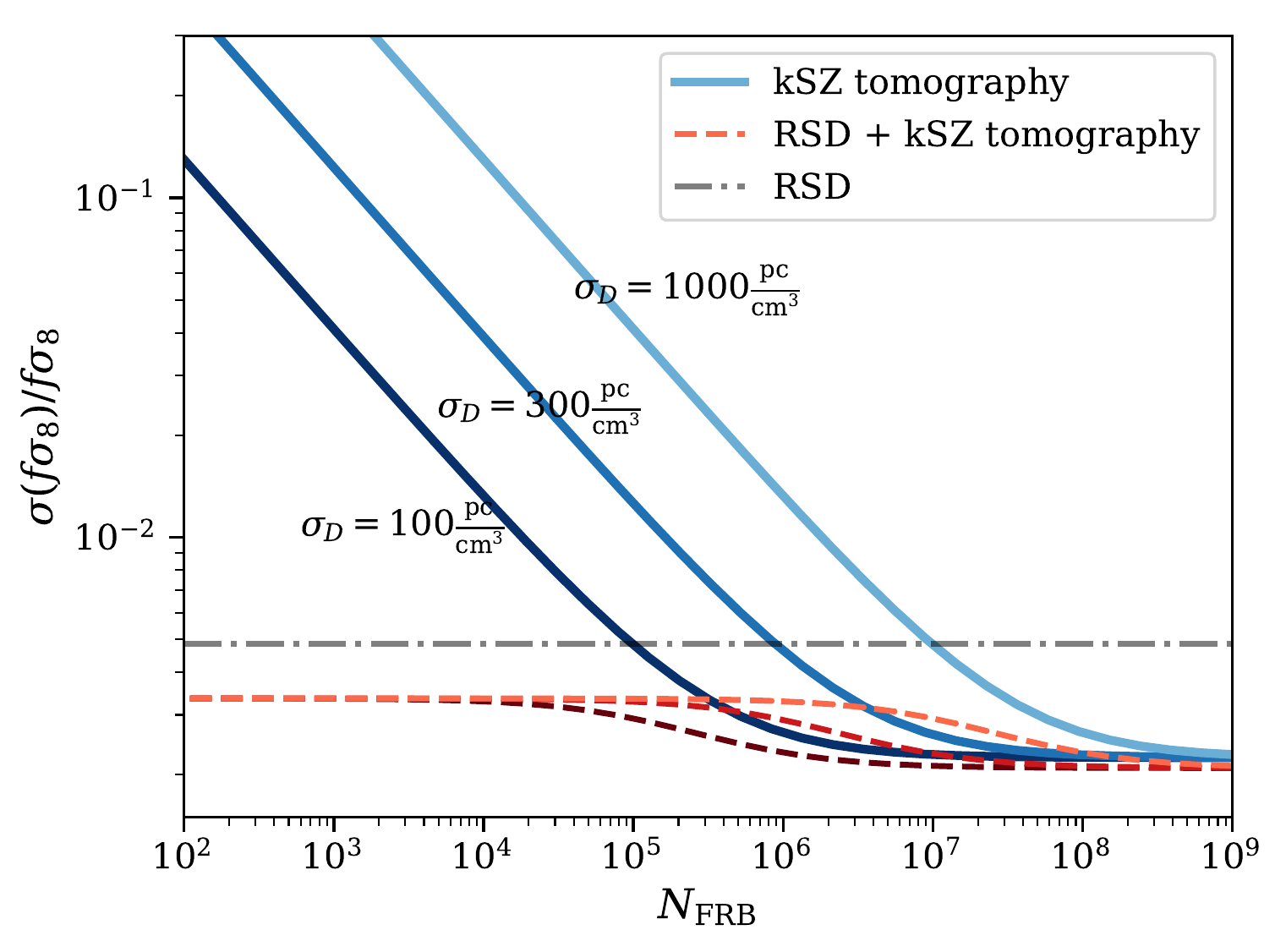}  
\caption{The uncertainty on the combination of cosmic growth rate and amplitude of matter fluctuations ${f\sigma_8}$ from kSZ tomography with CMB-S4 and DESI as a function of the number of FRBs, $N_{\rm{FRB}}$, available to break the `cluster optical depth degeneracy' through cross-correlation of FRB DMs with the same DESI galaxy sample. The blue lines show the constraint from kSZ tomography with various shades corresponding to choices of the uncertain RMS scatter of FRB DMs $\sigma_D$. If RSD information is used in conjunction with kSZ (red dashed lines), the degeneracy is already broken to some degree but further improvement is possible with FRBs. The grey dot-dashed line shows the constraint from DESI RSD alone.}
\label{fig:fs8}
\end{figure}

We consider survey combinations comprising of DESI and CMB-S4
and leave $N_{\rm{FRB}}$, which are localized with redshifts as a
free parameter. The assumed configurations of DESI and CMB-S4 can be found in
\cite{SmithKSZ2018}. DESI and CMB-S4 are used to obtain the above velocity
reconstruction. The reconstruction $\vrec$ can then be combined with the galaxy
overdensity field $\delta_g$ from DESI. The noise on the velocity reconstruction
is given by \cite{SmithKSZ2018},

\be
\begin{split}
N_{vv}(k) &=\mu^{-2} \frac{\chi_g^2}{K_g^2} \\
&\left[ \int \frac{k_S \, dk_S}{2\pi}  \left(\frac{P_{ge}(k_S)^2}{P_{gg}^{\rm tot}(k_S) C_l^{TT,\rm tot}}\right)_{l=k_S\chi_g} \right]^{-1},
\end{split}
\ee  
where $K_g$ is the kSZ radial weight function (defined in \cite{SmithKSZ2018})
at the galaxy shell redshift, $\chi_g$ is the comoving distance to the galaxy
shell redshift, and $C_l^{TT,\rm tot}$ is the total angular power spectrum of
CMB temperature anisotropies, including the late-time and reionization kSZ and foreground residuals after
multi-frequency cleaning.

This combination gives us the following power spectra

\begin{align}
P_{gg}(k,\mu) &= (b_g + f(z)\mu^2)^2P_{mm}(k), \label{eq:pgg} \\
P_{gv}(k,\mu) &= b_v \left(\frac{f(z)aH(z)}{k}\right)(b_g + f(z)\mu^2)P_{mm}(k), \label{eq:pgv}\\
P_{vv}(k,\mu) &= b_v^2\left(\frac{f(z)aH(z)}{k}\right)^2P_{mm}(k) \label{eq:pvv},
\end{align}
where $b_g$ is the linear galaxy bias, $P_{gv}$ is the galaxy-velocity cross
power spectrum, $P_{gg}$ is the galaxy auto power spectrum, and $P_{vv}$ is the velocity auto power spectrum.
We only include the redshift-space distortion (RSD) term $f\mu^2$ \citep{Kaiser1987} in Eqs.~\ref{eq:pgg} and~\ref{eq:pgv} if explicitly mentioned from here on. As mentioned in \cite{SmithKSZ2018}, the velocity reconstruction formalism explicitly shows how the `octopolar pair sum' estimator of \cite{Sugiyama:2016rue} that utilizes higher moments of the galaxy-velocity correlation in redshift space can break the optical depth degeneracy.  However, DMs from FRBs can be used as an independent way of breaking the optical depth degeneracy that is not affected by potential systematics in RSD measurements \cite{HirataRSD}. We thus do not include the $f\mu^2$ term in our baseline forecasts.

We can now forecast for cosmological parameters by constructing the Fisher matrix for the modes of the galaxy overdensity field and the reconstructed velocities

\be 
F_{ab} = \frac{V}{2}\int \frac{2\pi dkk^2}{(2\pi)^3} \int_{-1}^{1} d\mu \mathrm{Tr}\left[C_{,a}C^{-1}C_{,b}C^{-1}\right]
\ee
with covariance matrix,
\be C = 
\begin{bmatrix}
P_{gg}+N_{gg} & P_{gv} \\
P_{gv} & P_{vv}+N_{vv} 
\end{bmatrix} 
\ee
where $V$ is the total volume of the overlapping survey in $\rm Mpc^3$, $N_{gg} = 1/n_{\rm{gal}}$ is the shot noise contribution to the large scale galaxy power spectrum, with $n_{\rm{gal}}=1.7\times 10^{-4}~\rm{Mpc}^{-3}$ assumed for DESI.
We consider a cosmological model parameterized by the scale-independent growth
rate $f$ at $z=0$ and the amplitude of matter fluctuations $\sigma_8$ at
$z=0$. We perform a Fisher analysis for the parameterization
$\{b_g\sigma_8,f\sigma_8,b_v\}$ around fiducial parameters
$\{b_g=1.51,f=0.53,\sigma_8=0.83,b_v=1\}$ and use priors on $b_v$ obtained using
the results in Appendix B. We then obtain the marginalized constraint on
$f\sigma_8$ shown in Fig.~\ref{fig:fs8}.

\section{Results and Discussion}

We have shown that when the dispersion measures of FRBs are cross-correlated with a
galaxy survey, we can reconstruct the galaxy-electron power spectrum $P_{ge}$,
which is precisely the observable that breaks the kSZ optical depth degeneracy,
thus enabling cosmological applications of the kSZ effect. We find that the
cross-correlation of DMs from FRBs with a galaxy survey like DESI is detectable,
if around 100-1000 FRBs can be localized with sufficient redshift information to
place them behind the DESI sample (Fig.~\ref{fig:sn}). Such measurements translate into constraints on the optical depth of DESI galaxies at the 1\% level for 100,000 localized FRBs if 
$\sigma_D = 300$ $\rm{pc}/\rm{cm}^3$. In Fig.~\ref{fig:fs8}, we show how such
optical depth priors from FRB-galaxy cross-correlations translate to
cosmological growth rate measurements from kSZ tomography with CMB-S4 and DESI. We show that
$<1\%$ level constraints can be obtained with $N_{\rm{FRB}}>10^5$ and $\sigma_D
\sim 300$ $\rm{pc}/\rm{cm}^3$ independent of RSD measurements. Additionally, we
show improvements of up to 50\% can also be made when combined with RSD for very
large $N_{\rm{FRB}}$ values. These constraints saturate above $\sim 0.1\%$ due to sample variance in the distribution of electron fluctuations.

The numbers of FRBs considered here are significantly larger than the total number of detected FRBs to date, and no non-repeating FRBs have been localized to a host galaxy.  However, this dearth of observational data is set to change. The MeerKAT key project TRAPUM \citep{trapum2016} should localize $\sim 20$ FRBs/year in its coherent mode.  The deep synoptic array \citep{Ravi2018} should be able to localize $\sim$ 100 per year in its 200-dish phase. CHIME has already reported new FRBs and could find up to thousands per year.  While CHIME currently does not have localization capability, it could in principle be added. HIRAX expects to find 10-20 per day, with localization at the $\sim$ 30 mas level from southern African outriggers for a large fraction of those. Further in the future, SKA-MID should localize FRBs at 200 times the Parkes rate \citep{Macquart2015}, for $\sim 10^4$ per year.  So, getting to $10^5$ localized FRBs is feasible on a decade timescale with currently planned instruments. Since the FRB detection rate should scale like $A_{eff}^{1.5}$ for arrays with fixed primary beam, a factor of few increase in size on experiments like CHIME/HIRAX/DSA could feasibly push this up to $\sim 10^6$ events.

Beyond breaking the optical depth degeneracy,
the cross-correlation of DMs from FRBs with galaxy surveys provides constraints
on the baryon distribution in galaxies and clusters. On small scales, the shape
of $P_{ge} (k)$ is a measurement of the 1-halo electron free electron
profile. As previous theoretical works have shown for real space (not Fourier
space) \citep[e.g.,][]{McQuinn2014,Fujita2017,Ravi2018,ML2018}, this provides
valuable information on baryon density profiles of galaxies, groups, and
clusters. Additionally, the profiles inferred from FRB DMs are unbiased. Obtaining such profiles provides information on the impact of baryons on the matter power spectrum \citep[e.g.,][]{vD2011}, which is currently unconstrained by empirical measurements, but is extremely important for future cosmological measurements that aim to probe the matter power spectrum on small scales \citep[e.g.,][]{Semboloni2011,Eifler2015,Schneider2018}.

We have chosen the growth rate $f\sigma_8$ in these
forecasts as it is a model-independent parameterization for the physics probed
by cosmic velocities. The growth rate however can be affected by massive neutrinos, dark energy perturbations and modifications of General
Relativity. We thus expect that the breaking of the optical depth degeneracy
achieved using FRBs put forward in this work can yield significant constraints
on extensions of the standard model of cosmology. This will also require going
beyond the simplistic cosmological parameterization that we have considered here
(for example, incorporating marginalization over the Hubble constant and matter
density, while imposing priors from primary CMB measurements). These
explorations are left for future work.

\acknowledgments
We thank Jim Cordes, Colin Hill and Anze Slosar for comments on an early draft. We are
grateful to Matt Johnson and Moritz M{\"u}nchmeyer who helped develop the halo
model code used in this work. We also thank Ue-Li Pen and Vikram
Ravi for useful discussions. MSM is grateful to the Perimeter Institute for supporting his
visit during which this work was carried out.

\bibliography{msm}

\appendix
\section{Statistical error on $P_{ge}$ bandpowers}

The statistical error on a $C_l^{Dg}$ bandpower, defined by an $l$-range $[l_{\rm min}, {\rm lmax}]$,
can be derived as follows.  Working in the thin-shell approximation for simplicity, we take Eq.~(\ref{eq:thin_shell_snr})
for the total SNR, and restrict the $l$-integral to obtain the binned SNR:
\be
\mbox{SNR}^2_{\rm bin} = \Omega \int_{l_{\rm min}}^{l_{\rm max}} \frac{l\, dl}{2\pi} \frac{ (C_l^{Dg})^2 }{ (\mathcal{N}_l^{Dg})^2 }
\ee
The statistical error on the bandpower is then given by:
\ba
\Delta C_l^{Dg} &=& \frac{C_l^{Dg}}{\mbox{SNR}_{\rm bin}}  \nn \\
&=& \left( \Omega \int_{l_{\rm min}}^{l_{\rm max}} \frac{l\, dl}{2\pi} \frac{ 1 }{ (\mathcal{N}_l^{Dg})^2 } \right)^{-1/2}
\ea
In the thin-shell approximation, $C_l^{Dg}$ is related to $P_{ge}(k)$ by Eq.~(\ref{eq:cldg_limber}).
Therefore, we can recast the preceding result as the statistical error on a $P_{ge}$ bandpower over $k$-range $[k_{\rm min}, k_{\rm max}]$ to obtain Eq.~\ref{eq:delta_pge_frb}.

\section{Velocity bias prior}

At back-of-the-envelope level, the constraint on $b_v$ is $\sigma(b_v) = 1/\mbox{SNR}$, where
the SNR of the FRB-galaxy cross-correlation was given in Eq.~(\ref{eq:thin_shell_snr}).
However, this estimate is optimistic, since the SNR is obtained by summing all $k$-bins,
whereas the kSZ velocity bias only depends on $P_{ge}$ in a specific $k$-range.

To derive a better estimate for $\sigma(b_v)$ which we use in the rest of this work, we recall that the kSZ velocity-bias $b_v$ is defined by:
\be
b_v = \frac{\int dk_S \, F(k_S) P_{ge}^{\rm true}(k_S)}{\int dk_S \, F(k_S) P_{ge}^{\rm fid}(k_S)}  \label{eq:bv}
\ee
where
\be
F(k_S) = k_S \frac{P_{ge}^{\rm fid}(k_S)}{P_{gg}^{\rm tot}(k_S)} \left( \frac{1}{C_l^{TT,\rm tot}} \right)_{l=k_S\chi_g}
\ee
 and the integration range is over small-scale wavenumbers $0.1 $ Mpc$^{-1} < k < 10 $ Mpc$^{-1}$. We can obtain an estimate for the uncertainty $\sigma(b_v)$ on $b_v$ by relating it to the uncertainty $\Delta P_{ge}$ on the bandpowers of $P_{ge}$ through a quadrature sum of uncertainties, as is appropriate for uncorrelated bins that are normally distributed.  To do this, we define a large number of $k_S$-bins, with width $\Delta k_S$.
Replacing the integral in the numerator of Eq.~(\ref{eq:bv}) by a sum, the statistical error on $b_v$ is:
\be
\sigma(b_v)^2 = \frac{\sum F(k_S)^2 (\Delta P_{ge}(k_S))^2 (\Delta k_S)^2}{( \int dk_S \, F(k_S) P_{ge}^{\rm fid}(k_S) )^2}  \label{eq:sigma_bv_1}
\ee
where the sum in the numerator runs over $k_S$-bins.  For notational compactness, we rewrite Eq.~(\ref{eq:delta_pge_frb})
in the form:
\be
(\Delta P_{ge}(k_S))^{-2} = G(k_S) (\Delta k_S) \label{eq:dpge_compact}
\ee
where we have defined:
\begin{multline}
G(k_S) = \left( \frac{\chi_g}{n_{e0} (1+z_g)} \right)^{-2} \left( \frac{k_S
  \Omega}{2\pi} \right) \\ 
  \left( \frac{1}{(C_l^{gg} + 1/n_g^{2d})(C_l^{DD} + \sigma_D^2/n_f^{2d})} \right)_{l=k_S\chi_g}
\end{multline}
Plugging Eq.~(\ref{eq:dpge_compact}) into Eq.~(\ref{eq:sigma_bv_1}), we get our final expression for $\sigma(b_v)$:
\ba
\sigma(b_v)^2 &=& \frac{\sum F(k_S)^2 G(k_S)^{-1} (\Delta k_S)}{( \int dk_S \, F(k_S) P_{ge}^{\rm fid}(k_S) )^2} \nn \\
  &=& \frac{\int dk_S \, F(k_S)^2 G(k_S)^{-1}}{( \int dk_S \, F(k_S) P_{ge}^{\rm fid}(k_S) )^2}
  \label{eq:sig_bv}
\ea
where we have converted the sum back to an integral in the second line.  It is possible to
prove using this expression that $\sigma(b_v) \ge 1/\mbox{SNR}$, so our ``refined'' estimate for $\sigma(b_v)$
is more pessimistic than the back-of-the-envelope estimate $\sigma(b_v) \approx
1/\mbox{SNR}$, as anticipated. In this work, wherever a prior on $b_v$ is
assumed, the ``refined'' estimate derived here is used.

\end{document}